\documentclass[10pt]{article}
\pdfoutput=1
\usepackage{amsmath}
\usepackage{amssymb}

\usepackage{graphicx}

\usepackage{cite}
\usepackage{pdfsync}
\usepackage[usenames]{color}

\usepackage[labelfont=bf,labelsep=period,singlelinecheck=off,justification=justified]{caption}

\makeatletter
\renewcommand{\@biblabel}[1]{\quad#1.}
\makeatother

\date{}

\usepackage[hyperindex, colorlinks, hyperfootnotes = false, citecolor = black, breaklinks]{hyperref}
\usepackage{breakurl}

\begin{document}

\begin{flushleft}
{\Large
\textbf{Markov dynamics as a zooming lens for multiscale community detection: non clique-like communities and the field-of-view limit}
}
\\
\vspace*{.1in}
Michael T Schaub$^{1,2,\ast}$,
Jean-Charles Delvenne$^{3}$,
Sophia N Yaliraki$^{2}$ and
Mauricio Barahona$^{1,\ast}$
\\
\vspace*{.1in}

$^1$ Department of Mathematics, Imperial College London, London, United Kingdom
\\
$^2$ Department of Chemistry, Imperial College London, London, United Kingdom
\\
$^3$ Department of Applied Mathematics, Universit\'{e}  catholique de Louvain, Louvain-la-Neuve, Belgium
\\
$^\ast$ E-mail: michael.schaub09@imperial.ac.uk, m.barahona@imperial.ac.uk
\end{flushleft}

\section*{Abstract}
In recent years, there has been a surge of interest in community detection algorithms for complex networks. A variety of computational heuristics, some with a long history, have been proposed for the identification of communities or, alternatively, of good graph partitions. In most cases, the algorithms maximize a particular objective function, thereby finding the `right' split into communities. Although a thorough comparison of algorithms is still lacking, there has been an effort to design benchmarks, i.e., random graph models with known community structure against which algorithms can be evaluated. However, popular community detection methods and benchmarks normally assume an implicit notion of community based on clique-like subgraphs, a form of community structure that is not always characteristic of real networks. Specifically, networks that emerge from geometric constraints can have natural non clique-like substructures with large effective diameters, which can be interpreted as long-range communities.

In this work, we show that long-range communities escape detection by popular methods, which are blinded by a restricted `field-of-view' limit, an intrinsic upper scale on the communities they can detect. The field-of-view limit means that long-range communities tend to be overpartitioned. We show how by adopting a dynamical perspective towards community detection~\cite{Delvenne2010,Lambiotte2009}, in which the evolution of a Markov process on the graph is used as a zooming lens over the structure of the network at all scales, one can detect both clique- or non clique-like communities without imposing an upper scale to the detection. Consequently, the performance of algorithms on inherently low-diameter, clique-like benchmarks may not always be indicative of equally good results in real networks with local, sparser connectivity.
We illustrate our ideas with constructive examples and through the analysis of real-world networks from imaging, protein structures and the power grid, where a multiscale structure of non clique-like communities is revealed.


\section*{Introduction}

The analysis of community structure in complex networks has gained much attention in recent years and a variety of community detection algorithms have been proposed (for a recent overview see Ref.~\cite{Fortunato2010}). The reason for this interest is that by finding community structure in large networks one hopes to reveal relevant modules at mesoscopic scales that can affect or explain the global behavior of the system. Community detection may thus facilitate new insights into the structural and functional organization of a system (and the interplay between these two), as well as potentially serving as the basis for reduced descriptions of complex systems. Examples of important applications include networks from technological, physico-chemical, biological and medical data as well as data from the social sciences (see e.g. \cite{Fortunato2010,Castellano2009,Rives2003,Bullmore2009} and references therein).

Community detection algorithms are based on diverse notions of what makes a good community. Different mathematical and computational heuristics, some of them based on graph partitioning concepts, have been used to identify communities or to obtain an optimized split of the original network into smaller subgraphs with a community-like character. A common trait in many algorithms is to group nodes based on edge density: communities concentrate high edge weight within them, while having low edge weight between them. This structural notion has led to several heuristics including, among others, modularity~\cite{Newman2004,Newman2006a} and multiscale-Potts models \cite{Reichardt2004,Reichardt2006,Ronhovde2009} which have been coupled with different optimization algorithms (e.g., the Louvain method \cite{Blondel2008}) for the maximization of the corresponding cost functions. Recently, the Map equation framework has proposed an alternative notion that views communities as groupings of nodes that lead to concise descriptions (in an information-theoretical sense) for the process of communicating the position of a random walker within the network~\cite{Rosvall2007,Rosvall2008}.

To aid in the comparative evaluation of community detection algorithms,
there has also been an effort to design benchmark graph models with an embedded community structure~\cite{Newman2004,Danon2006,Lancichinetti2008,Lancichinetti2009a}.
However, in designing these benchmarks, a particular notion of community has to be adopted implicitly. As in many of the community detection algorithms described above, such benchmark models are based on the customary structural notion of community in terms of edge density. Therefore, the community structure is introduced as a \emph{stochastic clique-of-cliques},
i.e., as a hierarchy of interconnected realizations of Erd\"os-Renyi graphs with block-wise homogeneous, `all-to-all' edge densities.
This notion is motivated by the fact that many complex systems in the literature have been found to display small diameters and is indeed applicable to a wide range of networks, e.g. networks constructed from correlations or networks from the social sciences.

However, as we show below, there is a broad class of locally sparser networks where the assumption of `all-to-all' connectivity is not warranted yet may still contain relevant modular substructures, which we would like to identify via community detection. A community in this case corresponds to a set of nodes which have a stronger direct or \textit{indirect} connection with each other than with nodes outside their community. Such long-range substructures
cannot be modelled accurately by stochastic cliques.
This is the case in a variety of systems (e.g. biological and engineering networks) where entities are coupled in a complex manner via a chain of local interactions  such that not all entities of a module are directly connected, yet they are more strongly related to each other than to entities from a different subsystem.
Therefore, the evaluation of algorithms on benchmarks with clique-like communities might not be representative of their ability to identify relevant structures on real networks with localized, sparser connectivity where non-clique like communities may still exist.

The characteristics of communities in such networks (relatively sparse with a strong local structure) also highlight a limitation in popular community detection methods. It is already known that some of these methods (e.g., modularity) are affected by a  `resolution limit'~\cite{Fortunato2007}, a lower scale that establishes a minimum size below which communities cannot be detected. Infomap, on the other hand, seems to be immune to such a limit~\cite{Lancichinetti2009}. We show below that these methods also have a \textit{`field-of-view' limit}, an upper limit in the effective diameter of the communities they can detect. This field-of-view limit affects both modularity and Infomap.  Therefore, such structural methods contain an implicit scale and can only detect communities that lie within a range of effective sizes (dependent on each method and each graph) and might miss groups outside of it.

One way to correct for the field-of-view limit is to adopt an approach to community detection that intrinsically \textit{scans across scales}, such as the partition stability framework~\cite{Delvenne2010,Lambiotte2009}. Stability is a recently proposed dynamical approach to community detection which follows the time evolution of a Markov diffusion process on the graph incorporating increasingly longer paths on the network. This process entails a natural dynamic sweeping that can be understood as the application of a zooming lens to community detection.
It is important to remark that the sweeping process is inherent and key to the methodology: the dynamical basis implies a systematic tool by which the structure at all scales needs to be examined, without looking for the `right' scale. In doing so, stability is able to reveal communities without imposing a scale a priori or, indeed, community structure at multiple scales.
It has been shown~\cite{Delvenne2010} that in such a framework, the standard structural notion of community given by modularity is a particular case based on one-step transitions (i.e., detected at Markov time equal to 1). It can be shown~\cite{Schaub2011} that Infomap also considers one-step transitions in an averaged manner.

Under this interpretation, it is easy to understand why one-step methods (such as modularity and Infomap) cannot detect communities with large effective diameter, since these communities cannot be properly explored within one-step transitions. When faced with such communities, one-step algorithms tend to overpartition them. In fact, this feature provides us with an indicator that one-step methods are being applied to a community structure outside their range of applicability: when one-step methods return communities with large effective diameters, this can be seen as an indicator that the methods are operating on a graph which does not conform to their intrinsic notion of communities as cliques.   On the other hand, long range communities can be revealed at longer Markov times through the dynamic sweeping provided by stability.
We show below the relevance of these observations through the analysis of constructive examples and of real networks from biology, computer science and engineering.

\section*{Methods}

\subsection*{Notation}
Networks are here defined as undirected, connected, weighted graphs with $N$ nodes.  For simplicity, we consider non-bipartite graphs.  The connectivity of the graph is encoded by the weighted adjacency matrix $A$, a symmetric matrix where $A_{ij} = A_{ji} $ corresponds to the weight of the edge between nodes $i$ and $j$. The total weight of the edges is $m = \sum_{{i \geq j=1}}^{N} A_{ij}$.

\subsection*{Community detection methods: structural and dynamical interpretations}
The purpose of this section is to provide a dynamical re-interpretation of some popular community detection algorithms that are based on structural notions of community. It will be shown that such methods can be seen as one-step methods. When reinterpreted as one-step dynamical methods, it becomes possible  to understand the inherent assumptions or bias of structural methods towards the identification of short-range communities and the limits that this imposes on the detection of non clique-like communities.

\subsubsection*{A different interpretation of a well-known metric: Modularity as a one-step method}
The well-known modularity \cite{Newman2004,Newman2006,Newman2006a} has in recent years been used as one of the standard metrics to evaluate and optimize community structure. The original idea of modularity was intrinsically combinatorial:  it assigns nodes to communities such that the density of links inside a community is maximized, when compared to a random network with the same degree sequence.
It has been shown previously~\cite{Delvenne2010} that this measure can be interpreted dynamically as a one-step method.
Modularity, $Q$,  can be rewritten as~\cite{Delvenne2010}:
\begin{equation}
 Q = \text{trace }H^T\left[\Pi M  -\pi^T\pi\right]H,
\end{equation}
where $M = D^{-1}A$ is the one-step transition matrix of a discrete time random walk defined on the graph; $D$ is a diagonal matrix with the strengths of the nodes on its diagonal ($D_{ii} = d_i = \sum_{j} A_{ij}$); $\mathbf{\pi}_i = d_{i}/2m$ is the equilibrium distribution of the process; and $\Pi = \text{diag}(\pi)$. Furthermore the hard partitioning of the graph into $c$ communities is encoded into a $N \times c$ indicator matrix $H$ with $H_{ij} \in \{0,1\}$, where a 1 denotes that node $i$ belongs to community $j$.

Recasting modularity in this way suggests the following dynamical reinterpretation:
communities are assigned such that the overall probability of starting in a community and remaining there after one step is maximized compared to the probability of randomly ending up in the same community when at equilibrium. Hence modularity can effectively be seen as a one-step method and, as such, it tends to favor communities in which the Markov process spreads rapidly, such as clique-like groups with weak connections between them.  On the other hand, when the underlying communities are not clique-like (or have a large effective intra-community distance), modularity optimization can lead to overpartitioning, i.e., to the detection of artificially small communities. This is shown in the examples in the Results section.

\subsubsection*{Another one-step method: Infomap and clique-like communities}
Recently, Rosvall and Bergstrom \cite{Rosvall2008} have proposed the (Info-)Map framework for community detection.
The general idea behind the Map equation algorithm is to find a binary code with unique codewords for each node within a community that can be used to describe compactly the position of a random walker in the network.
In a network with a marked community structure, the probability of flipping membership is smaller. Therefore, one can compress the description by reusing short node names (codewords) within separate communities, similarly to the way street names may be used in
different cities throughout a country~\cite{Rosvall2008}.

Although different in spirit to modularity, it is interesting to note that the Map equation, much like modularity, is also based solely on the equilibrium distribution $\pi$ and a one-step process that reflects inter-community transitions in an block-averaged manner. Indeed, it can be shown that the Map equation formalism does not distinguish different connectivity structures inside communities~\cite{Schaub2011}. This reflects an implicit assumption about fast mixing communities and, consequently, a homogeneous `all-to-all' connection pattern (i.e., clique-like communities in the sense of having short effective diameter) matches best the inherent notion of community of the algorithm.
Hence, Infomap is prone to finding communities that can be well approximated as stochastic cliques. Indeed, Infomap performed as one of the best algorithms in a recently conducted benchmark test where communities are defined as stochastic cliques~\cite{Lancichinetti2009}. This also explains why, unlike modularity, the Map equation is not known to be affected by the resolution limit: the fact Infomap is greedy with respect to locally dense substructures makes it well suited for detection even at the finest scales. However, this desirable, designed-for feature also means that when the graph under  consideration has slowly mixing communities, Infomap displays a pronounced overpartitioning effect, even more so than modularity.

\subsubsection*{Stability as a dynamical framework for community detection: sweeping across scales}

Communities with large effective diameters, e.g., non clique-like communities, can be missed by the methods above because they may not be discernible with one-step measures.
This raises the question of how to detect non clique-like communities?  Intuitively, a means to account for their structure is to consider walks of lengths greater than one. This can be done in a principled way using the recently proposed partition stability framework~\cite{Delvenne2010,Lambiotte2009}.
The idea underpinning stability is to define a Markov (diffusion) process on the graph and follow how the probability flow spreads out over time.
Rather than looking at one-step measures,
stability takes into account walks of increasing length systematically by looking at larger times and, in doing so, it can reveal communities at different scales: in general, the longer the time, the coarser the partition. Alternatively, one can think of the Markov time as providing a \textit{zooming lens} that scans the structure of the graph from the finer to the coarser structure.
Importantly, stability has been shown to provide a unifying framework for several measures of community detection including modularity and spectral partitioning~\cite{Delvenne2010}.
In the following, a brief introduction to the stability measure will be given. For a more detailed exposition see \cite{Delvenne2010,Lambiotte2009}.

The stability of the partitioning of a graph can be defined via the autocovariance of a Markov process taking place on the graph. The definition can be based on both discrete and continuous times.  Here we consider the continuous-time Markov process on the graph governed by the following (Laplacian) dynamics:
\begin{equation}
 \mathbf{\dot p} =
 - \mathbf{p}\;[D^{-1}L],
\end{equation}
where $\mathbf{p}$ is a $1\times N$ probability vector, $D$ is the diagonal matrix with the strengths $d_i$, and $L = D- A$ is the graph (combinatorial) Laplacian matrix.
Under the assumptions made above (undirected, connected, non-bipartite graphs), the stationary distribution $\pi$ of this dynamics is
$\mathbf{\pi}_i = D_{ii}/2m$.
Now we can define the clustered autocovariance matrix of the graph at time $t$ as:
\begin{equation}
 R_t = H^T\left[\Pi \exp(-t D^{-1}L) -\pi^T\pi\right]H,
\end{equation}
with $\Pi = \text{diag}(\pi)$ and the matrix $H$ encoding the partitioning as defined for modularity above.
Note that each entry $[R_t]_{ij}$ is the probability for a random walker to start in community $i$ at stationarity and after time $t$ end up in community $j$ minus the probability of two independent walkers to be in $i$ and $j$ at stationarity.

We can then write the \emph{stability of a partition} $H$ at time $t$ as:
\begin{equation}
 r(t,H) = \text{trace } R_t.
\end{equation}
This is a global quality function for a given graph and partition that changes as a function of time.  It is easy to see that modularity is a particular case of the linearization of the stability (or of its discrete analog) at time $t=1$~\cite{Lambiotte2009}.
The stability $r(t,H)$ can now be optimized in the space of partitions $H$ with any optimization method for graph clustering. In the present work, this has been done with the efficient Louvain method~\cite{Blondel2008}.
The effect of time is intuitive: with increasing time the Markov process explores larger regions of the graph, so the Markov time acts as a resolution parameter that enables us to identify community structure at different scales.  Communities are identified as subgraphs within which the probability distribution of the process is more contained over a time $t$  than otherwise expected at stationarity.
Importantly, stability does not aim to find \textit{the} best partition, but rather tries to reveal relevant clusterings at different scales through the zooming process that occurs naturally through the dynamics. A relevant partition should be both persistent over a comparably long timescale and robust with respect to slight variations in the graph structure and/or the optimization~\cite{Lambiotte2010, Delmotte2011}. In order to quantify the robustness of the partitions, we use the variation of information to measure the similarity between partitions~\cite{Meila2007,Delmotte2011}.

Stability differs from one-step methods in its intrinsic multistep character, as it is based on the exponential of the full adjacency matrix. Furthermore, stability does not introduce any effective assumption towards a block-averaged transition matrix and is thus not biased towards a particular structural model of clique-like communities. The fact that stability uses the diffusion dynamics from each node taking into account walks of all possible lengths, rather then looking at one-step transitions only, has an important bearing on the detection of non clique-like communities.
As the Markov time increases, stability is able to find the cohesion within non clique-like communities as the multi-step density between nodes in such a community is increased.
The dynamic zooming provided by the Markov process is the key characteristic of the stability framework. Such an approach does not focus on a particular scale but rather provides a means to establish the presence of robust partitions that can appear at any given scale or, indeed, reveal the existence of a multiscale community structure. If no communities are present, e.g. in pure Erd\"os-Renyi random graphs, stability returns the absence of robust communities at any scale. The effect of the Markov dynamics as a zooming lens, scanning from finer to coarser resolution, provides a means to ameliorate the effect of the resolution and field-of-view limits. We should remark that the Markov time does not impose a dynamic on the network necessarily but can rather be seen as a device to reveal the potential community structure in the network (even if the network does not have an intrinsic dynamics).

\subsection*{Community structure in benchmark graphs and in different applications}

A number of benchmark graph models have been proposed for the evaluation of community detection algorithms (for some widely used examples see~\cite{Newman2004,Danon2006,Lancichinetti2008,Lancichinetti2009a}).
Nearly all of these benchmarks share a particular notion of community: typically a parameter controls the ratio of the probability for a node to connect within its own community versus the probability to connect to a node outside the community.
Some more recent models~\cite{Karrer2011} allow for more heterogeneous degree distributions of the nodes, yet a community can basically still be described as a \emph{homogeneous} substructure and
the community structure may be thought of as a probabilistic realization of a `clique of cliques':
\begin{equation}
 p(i,j) \propto k_ik_j f(C_i,C_j),
\end{equation}
where $k_i,k_j$ are the degrees of node $i$ and $j$ respectively and $f$ is a function of the associated communities $C_i,C_j$ of $i$ and $j$ alone.
Therefore, these random graph models can be thought of as `mean field' or `block models', in that intra- and inter-community structure is approximated by average connection properties.

If we think of Markov diffusion processes or flows taking place on these graphs, such networks exhibit fast exploration of all nodes within each community---due to the clique-like structure of the community any node in the community is reached in one step. Although graphs of this type are indeed found in real applications (such as networks constructed from correlation measurements or in some instance of social groupings), a wide spectrum of real networks are not of this form because they do not display an `all-to-all' connection pattern.
An important example of networks that cannot support clique-like connection patterns is that of geographically embedded networks, such as power grids, sensor-networks, river-networks, road- and train-networks and other transport and supply and distribution networks.
Similarly, networks containing constraints, in some cases dictated by geometry or by some other cost functions, such as higher-dimensional grid or lattice-like structures originating from physical and biological systems, will not display homogeneous block-like structures in their connectivity patterns. We show some examples of these networks in the Results section below.

This issue is generic in the sense that it applies to graphs that have an intrinsic geometric or constraint-driven structure even if it emerges from an abstract data structure. Examples include graphs with a low doubling dimension~\cite{Gupta2003} which are related to graphs that can be naturally embedded in a low-dimensional Euclidean space. Indeed, it has been conjectured that the Internet autonomous systems network--a prominent example of a network pegged to geographical constraints--has low doubling dimension~\cite{Fraigniaud2008} and can be embedded naturally into a two-dimensional hyperbolic plane~\cite{Boguna2010}.
Other networks with a similar structure emerge from the projection of  high-dimensional data onto a lower dimension while preserving the local structure of the data~\cite{Tenenbaum2000} or in web graphs with a tendril-like structure in the periphery.
In all these networks, the connection probability between nodes is strongly influenced by locality and it tends to be sparser and inhomogeneously distributed within and in between different structural components. Invoking again a diffusion process, these graphs may contain non clique-like communities within which diffusion is slow, such that they cannot be identified with standard structural, one-step community detection methods.
The fact that a wide range of real networks are not of a mean-field type and their possible community structure may not be clique-like should be taken into account when comparing community detection algorithms on benchmarks which have been developed with the implicit assumption of clique-like communities.

\section*{Results}
We now analyze a series of constructive and real-world examples in which communities are significantly different from clique-like structures. For all the results presented below, we run the respective community detection algorithms 100 times with a different random seed and select the best partition found as the community structure. For both modularity and stability, the optimization has been performed with the Louvain algorithm~\cite{Blondel2008}. Infomap has its own implicit optimization method. To assess the robustness of communities obtained by the stability method,  we used the variation of information~\cite{Meila2007} of all solutions found by the optimization.

\subsection*{Constructive examples: Non clique-like communities with low intra-community diffusivity}

\begin{figure}[!ht]
 \centering
\includegraphics[width=1\columnwidth]{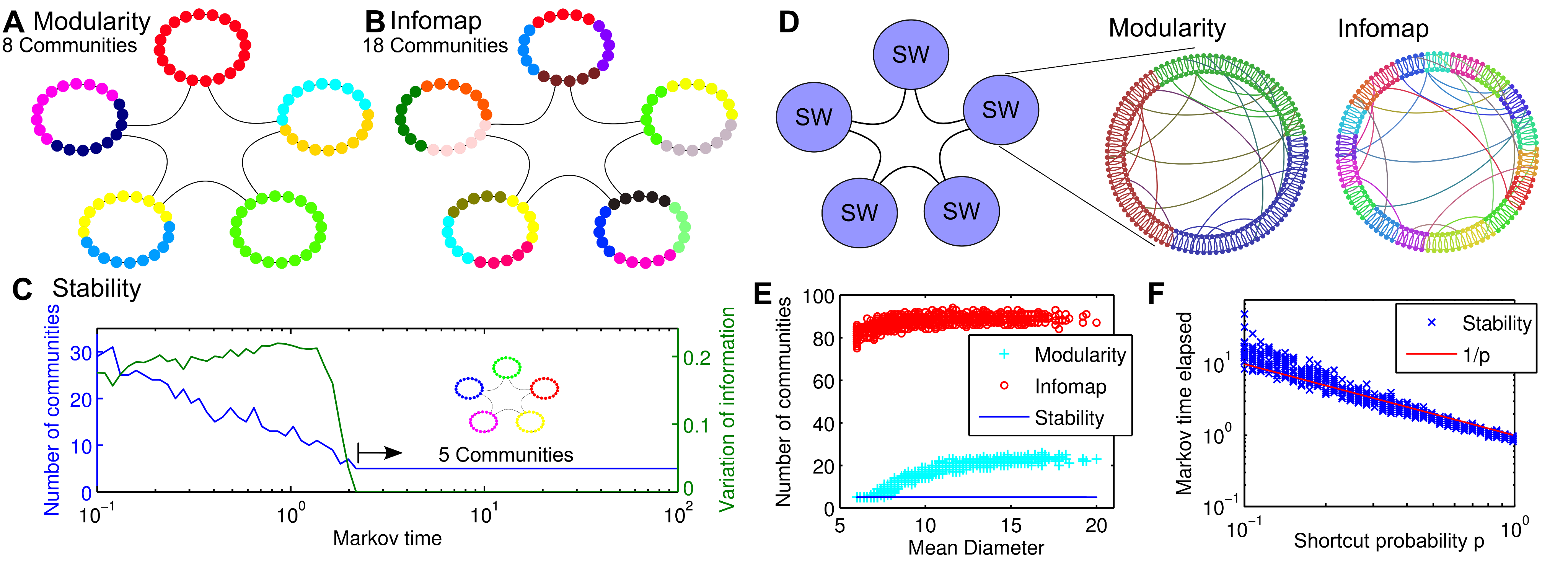}
 \caption{\textbf{Constructive networks with non clique-like community structure}. \textbf{A-C} Ring of rings: 5 rings of 20 nodes in a ring configuration. The edges within each ring are 5 times stronger than between rings.  Optimal communities according to: \textbf{A} modularity (8 communities found) and \textbf{B} Infomap (18 communities found). \textbf{C} Analysis with stability: number of communities found (blue) and average variation of information (VI) of the partitions found (green) as a function of Markov time. The average VI is obtained from 100 runs of the Louvain algorithm.
 Starting at $t \approx 1.96$ the correct partitioning into 5 communities is detected as a persistent, robust partition with vanishing VI. No other stable partition is detected at any other Markov time (as shown by the high values of VI).  \textbf{D-F} Ring of small worlds: 5 SWs of 200 nodes in a ring configuration. The edges inside each SW are 5 times stronger than those between different SWs. The SW property, and thus the diameter of the SWs, is varied by varying the shortcut probability $p\in[0.1,1]$ (see text for details). \textbf{D} Examples of partitions obtained by modularity and Infomap for $p=0.1$. \textbf{E} Number of communities detected as a function of the measured mean diameter of the small world subgraphs for Infomap, modularity and stability: Infomap never detects the SWs; modularity detects the SWs only when they have short enough diameter; stability always detects the SWs (at different Markov times) as the only stable partition. \textbf{F} Markov time elapsed until the SWs are detected by stability as a function of the shortcut probability.}
 \label{fig:1}
\end{figure}

Before considering real-world networks,  we illustrate our ideas with constructive toy examples to exemplify the notion of non clique-like communities. The first example is a  `ring of rings', in which 5 rings of 20 nodes with \textit{strong intra-ring} edges are linked via weak edges to each other (Fig.~\ref{fig:1}A-C). Given our discussion above, it would be desirable that community detection algorithms should reveal the strongly linked ring communities. Indeed, the strong rings correspond to the notion of community as the equivalent of the connected components when the graph is `almost disconnected', i.e., in the limit when the weaker edges become non-existent.
However, as shown in Figure \ref{fig:1}A-B,  one-step approaches, such as modularity and Infomap, fail to recover these communities and return `optimal' clusterings that are overpartitioned with many communities of no individual relevance.

Following our discussion in the Methods section, these results can be understood as a consequence of the locally greedy, one-step characteristics of these algorithms. The Infomap algorithm obtains 18 communities as this algorithm does not have any incentive to create communities that go beyond small ring segments, as such communities have low one-step escape probability while being as locally clique-like as possible. In fact, Infomap fails to recognize the rings even when all edges in the graph (within and between communities) have equal weights.
Modularity also optimizes according to one-step transitions and, in the case of the rings, such a short horizon is not enough to identify the rings of length 20 and it returns a partition into 8 communities. This effect becomes more acute when the length of the rings increases, e.g.,  with 2 rings of 50 nodes modularity finds 10 communities instead of the expected 2.

In contrast, the stability formalism identifies only one persistent and robust partition: the right split into 5 rings (Fig.~\ref{fig:1}C).
Stability detects the communities when the dynamic zooming reaches sufficiently long Markov times. In this case, this occurs at a Markov time greater than 1, thus explaining why modularity is unsuccessful. Indeed, above $t \simeq 2$,  a long-lasting plateau of stability corresponding to 5 communities emerges and this partition is robust as given by the vanishing value of the variation of information calculated for 100 optimizations of the stability at each time point.

As explained above, the Markov time at which the structure is detected is indicative of an effective intra-community diffusion distance that needs to be spanned to make the community identifiable. This is akin to a measure of the diameter of the community (specifically in the case of quasi-regular graphs).  The fact that both Infomap and modularity are effectively one-step methods means that if the communities they detect are the `right ones', they should always have low diameter. Conversely, when the communities returned as optimal by Infomap and modularity have large diameters, this can be seen as an indication of an overpartitioning effect by these methods. In that case, both algorithms are operating in a regime for which they were not designed since they will always try to find locally dense or one-step structures and the communities may thus be too fine grained.

This observation can be used to provide a check for the appropriateness of those methods for the particular network analyzed. As a simple proxy, one can monitor the diameter of the detected communities\footnote{The diameter is just an easily computable indicator of this effect but other more appropriate measures could be used, specially in the case of non-regular graphs. Here we calculate the average of the maximum of the shortest paths in the subgraphs induced by the partioning.}.
For instance, in the case of the ring of rings, the corresponding average diameters of the communities found by each method are $\Delta_{\text{Mod}} \approx 14$ and $\Delta_{\text{Map}} \approx 4.56$, indicating a potential overpartitioning.  Similar diameters are observed in the real-world examples shown below, as does the fact that the communities detected by Infomap tend to be smaller in diameter than the ones found by modularity, a reflection of Infomap being more greedy towards locally clique-like structure. This feature of Infomap, which makes it immune to the resolution limit, makes Infomap more sensitive to the field-of-view limit, and hence to overpartitioning.

Further insight into the importance of the intra-community distance in non clique-like communities can be gained by considering another constructive example: a ring of 5 small-world (SW) graphs~\cite{Watts1998} with 200 nodes each in which edges inside SWs are five times stronger than edges between SWs (Fig.~\ref{fig:1}D-F). The SWs are constructed using the CONTEST toolbox~\cite{Taylor2009} as follows: start from a ring with nearest and next-nearest connections and add a random shortcut to each node with probability $p$.
As shown in Figure~\ref{fig:1}D, both Infomap and modularity suffer from severe overpartitioning: when the SW graphs have few shortcuts, Infomap finds 87 communities and modularity returns 27 communities while stability finds the right split into 5 communities.

As the shortcut probability is increased, the diameter of the SW is reduced while their local structure is basically unaffected going from a `large world' to a `small world' (this is the classic finding by  Watts and Strogatz~\cite{Watts1998}). From our viewpoint, this means that the SWs become smaller (i.e., their diameter decreases) and these communities should become easier to detect by one-step methods. Indeed, it is shown in Figure~\ref{fig:1}E that modularity detects the correct number of communities when their mean diameter falls below 7. On the other hand, Infomap does not detect the SWs as communities in the range shown in Fig.~\ref{fig:1}E, and only detects the SWs when the diameters fall below a diameter of around 4 (not shown). Again, this highlights the bias of Infomap towards locally clique-like communities with short effective diameters. The same feature that makes Infomap successful in dealing with the resolution limit appears here at the basis of its susceptibility to the field-of-view limit.

Figure~\ref{fig:1}E also shows that stability consistently detects the right community structure with 5 SWs as the only persistent and robust partition, for all densities of shortcuts, i.e., no matter how small or large their diameter is. According to our dynamic viewpoint, it is expected that as the diameter decreases, the SW communities will be detected at smaller Markov times. This is presented in Figure~\ref{fig:1}F, where we show that the Markov time at which the SWs become identifiable by stability decays roughly as the inverse of the shortcut probability. Satisfyingly, this correlates well with the dependence of the mixing time in SWs estimated from the spectral gap, i.e., the first non-zero eigenvalue of the Laplacian~\cite{Barahona2002}.

\subsection*{Networks with non clique-like communities from diverse real-world applications}

As shown above, the key limitation of modularity and Infomap for the analysis of non-clique community structures is their reliance on a single-step (structural) notion of community, which leads to existence of a fixed scale in the algorithm. In contrast,  stability zooms across scales and, importantly, the scanning through all scales is an essential feature of the algorithm.
By scanning through time and considering all relevant features detected along the way, a multi-scale structure can be revealed, yet without imposing a scale \textit{a priori}. In cases where there is no community structure, as for Erd\"os-Renyi random graphs, the algorithm does not find structure at any scale. If there is one or more scales, they can be found through the sweeping.
Therefore, in stability, the potentially multi-scale structure of the graph is explored through a multi-step process given by the Markov time.

%

\paragraph{Graphs from image analysis}
As a first real application we consider an image segmentation problem in which the aim is to analyze an image and find meaningful substructures without \textit{a priori} knowledge or guidance.
One approach towards this problem is to create a graph representation of the image, in which each node corresponds to a pixel in the original image and the weight of the edges is computed according to image properties such as distance between the pixels, difference in intensity and/or color, etc.~\cite{Browet2011}. The resulting graph can then be analyzed for community structure as a means to detect meaningful substructures in the image. Clearly, the graphs thus generated will not have homogeneous, clique-like community structure, since the graph is generated from an image with a two-dimensional structure and it incorporates a distance metric.

\begin{figure*}[!ht]
 \centering
  \includegraphics[width=1\columnwidth]{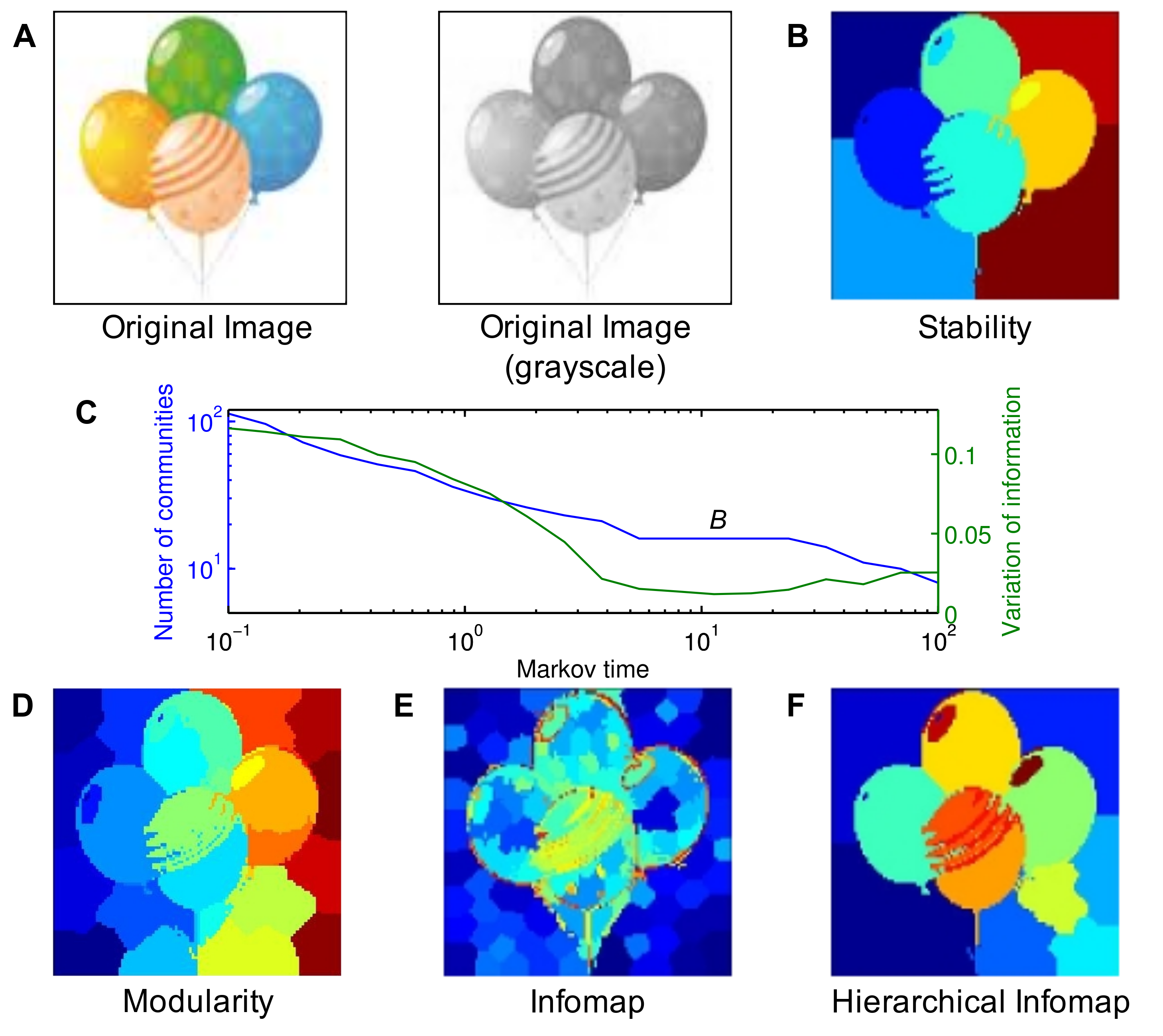}
 \caption{\textbf{Image segmentation via community detection}. \textbf{A} Original image in color and the grayscale version used in the analysis. \textbf{B} Image segmentation (false color plot) found with stability (16 communities). \textbf{C} Stability analysis of the image graph. 
The partition in \textbf{B} corresponds to the plateau at Markov time $t=11.3$, where a minimum of the VI also occurs. \textbf{D-F} Image segmentation (false color plot) obtained from different community detection methods: \textbf{D} modularity (37 communities), \textbf{E} Infomap (213 communities), \textbf{F} hierarchical Infomap (15 communities; highest hierarchical level).}
 \label{fig:2}
\end{figure*}

In Figure~\ref{fig:2} we present the results of the analysis of a sample image~\footnote{The image is freely available in various sizes in png format at \url{http://www.iconarchive.com/show/christmas-icons-by-mohsenfakharian/balloons-icon.html}. For the present analysis, a jpg variant of slightly different size has been used.} of size $102 \times 102$ pixel. The graph we analyze is constructed from the grayscale version in Fig.~\ref{fig:2}A. As expected from our discussion above, Figure~\ref{fig:2}D-E shows that both Infomap and modularity lead to an over-segmentation of the image into 213 and 37 communities, respectively, with average diameters larger than one: $\Delta_{\text{Map}} \approx 3.67$ and $\Delta_{\text{Mod}} \approx 8.27$, a fact that is indicative of non clique-like communities. On the other hand, the dynamic zooming provided by stability finds a naturally robust partition into 16 communities, which emerges at a Markov time of around 11 (Fig.~\ref{fig:2}B-C) and corresponds well with the underlying features of the image. The robust partition is indicated both by its persistence in terms of Markov time and by its robustness, as indicated by a minimum in the variation of information between the partitions found by the optimization.
For comparison, we also evaluated in Fig.~\ref{fig:2}F hierarchical Infomap~\cite{Rosvall2011}, which has been proposed recently and allows to agglomerate hierarchically the communities obtained by the Map equation formalism into different levels.  Our analysis shows that only the clusterings obtained at the highest level of hierarchical Infomap provide a perceptual improvement and the hierarchical scheme still leads to communities split into subcomponents of no obvious significance. This example from image segmentation highlights the relevance and significance of a notion of community that deviates from the usual clique-like assumptions, which can only be detected with multi-step community detection algorithms, such as stability optimization.
%

\begin{figure*}[ht!]
 \centering
\includegraphics[width=1\columnwidth]{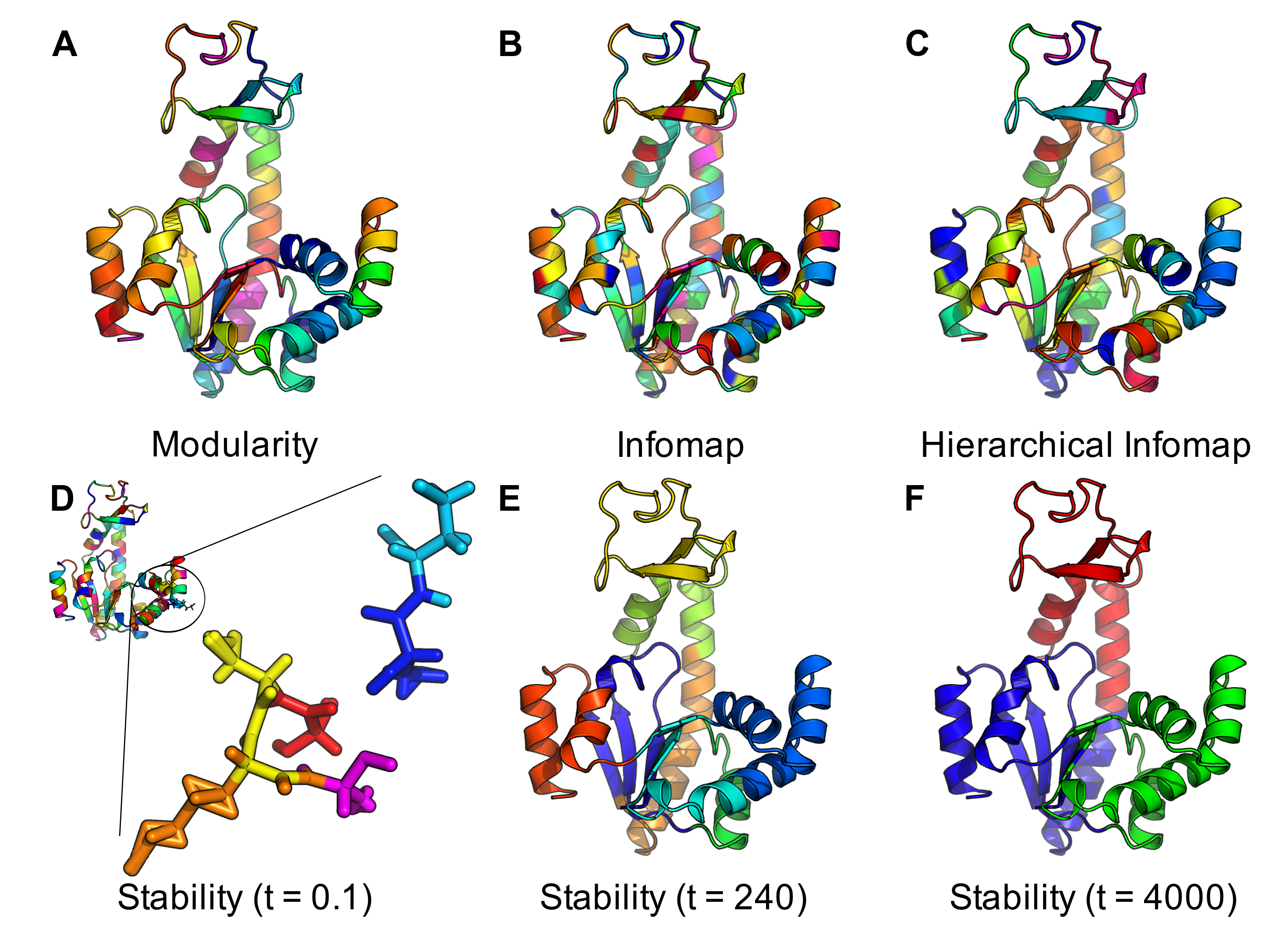}
 \caption{\textbf{Analysis of the structural graph of the protein Adenylate Kinase (AdK)}. \textbf{A-E} Visualization of the communities found by the different algorithms (adjacent regions in the same color correspond to communities): \textbf{A} Modularity (69 communities); \textbf{B} Infomap (421 communities); \textbf{C} hierarchical Infomap (58 communities). \textbf{D-F} Some of the robust communities found by stability at different Markov times: \textbf{D} $t=0.1$ (206 communities), the communities capture the amino acids of the protein (214 amino acids);
 \textbf{E} $t=240$ (8 communities), the communities correspond approximately to secondary structure of the protein (c.f. \cite{Delmotte2011});
 \textbf{F} $t=4000$ (3 communities), the communities correspond to the functional domains of the protein that operate at slow timescales.
 Note that stability finds meaningful substructures also for other times not shown~\cite{Delvenne2010,Delmotte2011}.}
 \label{fig:3}
\end{figure*}

\paragraph{Protein structure analysis}
Proteins are a class of macromolecules with complex three dimensional spatial structure which exhibit a hierarchy of motions that are intimately coupled to their function. Structural analysis that can shed light into their function is a very active area of research. Although it is well known that identifiable motifs appear at different time and length scales, a coherent methodology that can provide an integrated description of the hierarchy of structures from the bottom-up remains elusive.
As a second example, we show in Figure~\ref{fig:3} the analysis of the protein Adenylate Kinase (AdK), an enzyme which functions by performing an opening and closing global motion at slow timescales. Therefore, AdK has been studied both experimentally and theoretically  as a model for the analysis of hierarchical dynamics of proteins and as a benchmark for method development (see, for example~\cite{Delmotte2011} and references therein).
In this case, the graph is created from structural data, i.e., from the positions of atoms in three-dimensional space and the interactions between them. This results in a graph in which the nodes are atoms and the edges correspond to bonds and chemical constraints (for details see \cite{Delmotte2011}).

The structural and geometric origin of the graph leads to a non clique-like community structure. Again, this causes overpartitioning  for both Infomap and modularity, as can be seen in Figure \ref{fig:3}A-B: Infomap returns 421 communities with average diameter $\Delta_\text{Map} \approx 4.05$ while modularity detects 69 communities with average diameter $\Delta_\text{Mod} \approx 10.12$. The large diameters again indicate that both methods are operating in a regime that does not match their intrinsic assumption of what constitutes a good community. Indeed, the partitions obtained by these methods do not reflect the dynamical and structural features that are prominent in AdK~\cite{Delvenne2010,Delmotte2011} and even the highest level in hierarchical Infomap (Fig.~\ref{fig:3}C) is overpartitioned and does not provide an appropriate coarse graining in this case.
On the other hand, the robust partitions detected by stability disclose the multiscale structure of the protein graph, revealing important functional and structural subunits, such as amino acids, secondary structures and conformational substructures as exemplified in Figure \ref{fig:3}D-F~\cite{Delvenne2010,Delmotte2011}.

\paragraph{Power grid network}
Our final example is the analysis of a classical technological network, namely the power grid of Continental Europe~\footnote{Dataset taken from \url{http://www.termoenergetica.upc.edu/marti/index.htm}}. This network is based on data from the Union for the Coordination of Transmission Energy (UCTE) and has been analyzed previously for robustness to targeted attacks~\cite{Rosas-Casals2007,Sole2008}.
As this network is embedded geographically and is constrained by engineering costs, its structural properties are far from `all-to-all' connectivity and we expect non clique-like communities.

\begin{figure*}[!ht]
 \centering
\includegraphics[width=0.5\columnwidth]{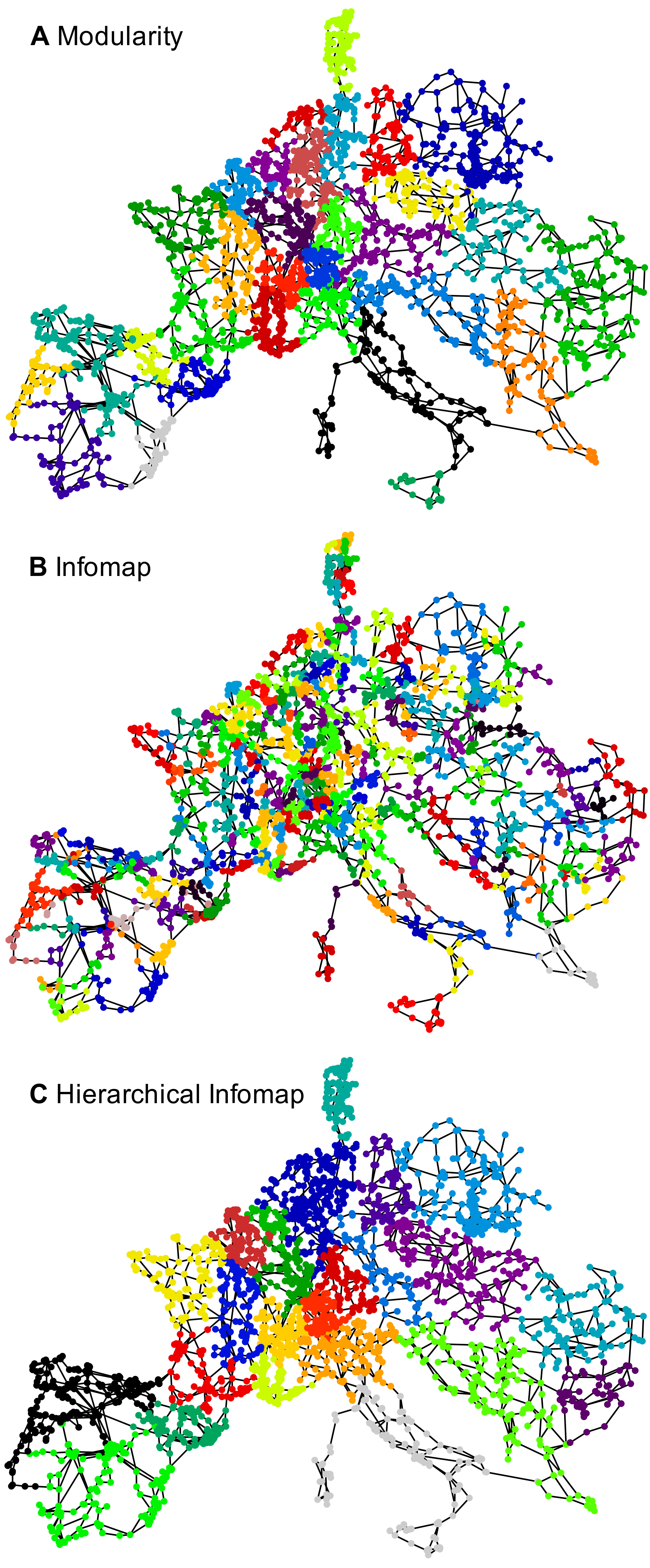}
 \caption{\textbf{Community structure analysis of the European power grid with different one-step methods}: \textbf{A} modularity (32 communities); \textbf{B} Infomap (254 communities); \textbf{C} hierarchical Infomap (24 communities; top level of hierarchy).
}
 \label{fig:4}
\end{figure*}

\begin{figure*}[!ht]
 \centering
\includegraphics[width=\columnwidth]{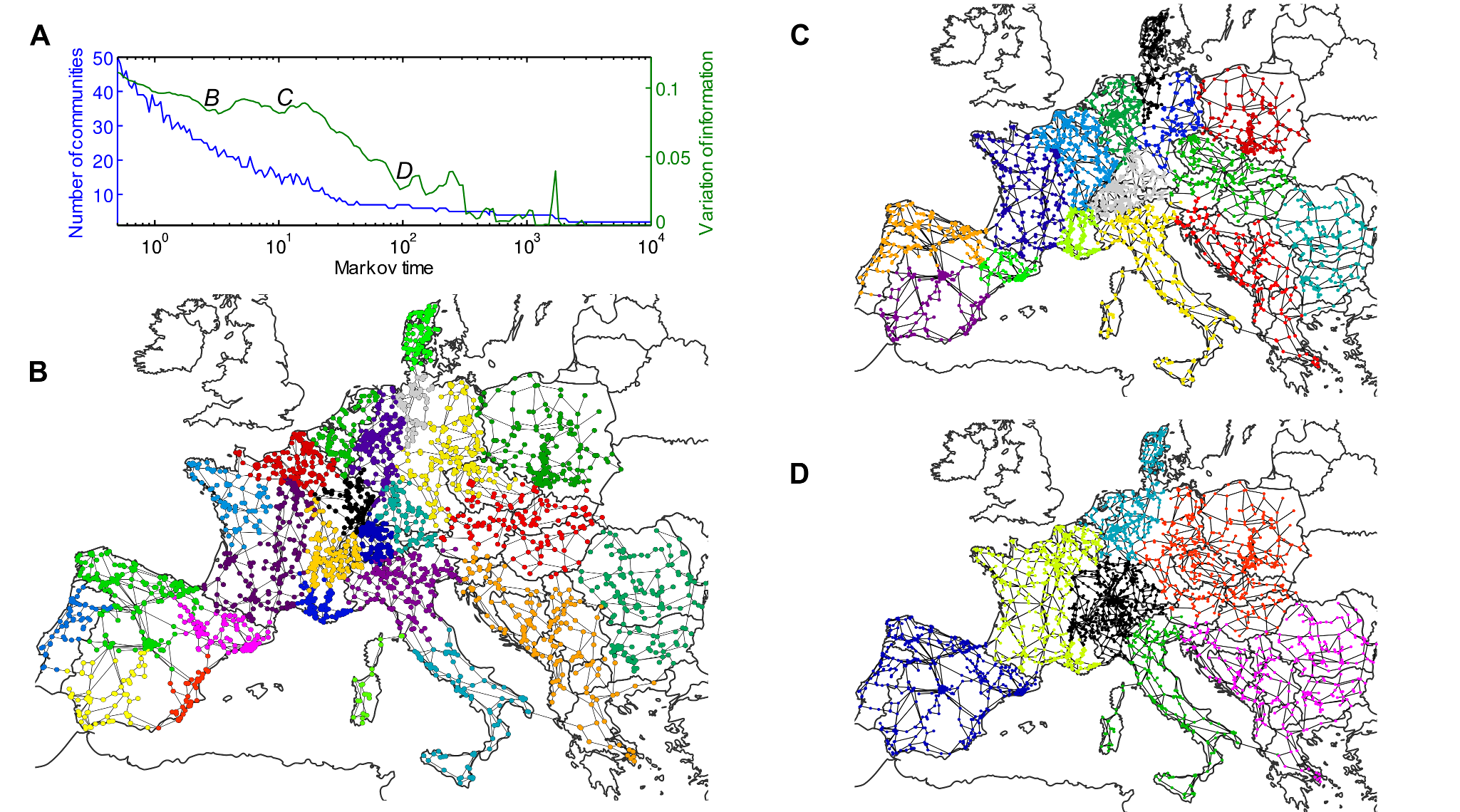}
 \caption{\textbf{Community structure analysis of the European power grid with stability}. \textbf{A} Stability analysis of the community structure of the power grid graph: number of communities found (blue) and average VI (green) vs Markov time. In this case, 1000 initializations of the Louvain algorithm have been used to find the best community and compute the variation of information. \textbf{B-D} Stability finds robust partitions for different Markov times that seem to be related to known structure in the power grid.
The partitions shown  correspond to Markov times: \textbf{B} $t = 2.63$ (25 communities), \textbf{C}  $t = 11.76$ (15 communities) and \textbf{D} $t=94.79$ (7 communities).}
 \label{fig:5}
\end{figure*}

The community structures obtained by stability, modularity and Infomap
 are shown in Figures \ref{fig:4}~and~\ref{fig:5}.
The results for Infomap and modularity follow the same pattern as above: Infomap overpartitions into 254 communities with average diameter $\Delta_\text{Map} \approx 4.91$ while modularity returns 32 communities with $\Delta_\text{Mod} \approx 13.84$ (Fig.~\ref{fig:4}A-B). Both methods result in a fractured representation of the European power grid. Hierarchical Infomap finds non-meaningful partitions for low hierarchical levels and improved partitions only at its highest level, although still displaying segregated small regions.

\begin{figure*}[!ht]
 \centering
\includegraphics[width=\columnwidth]{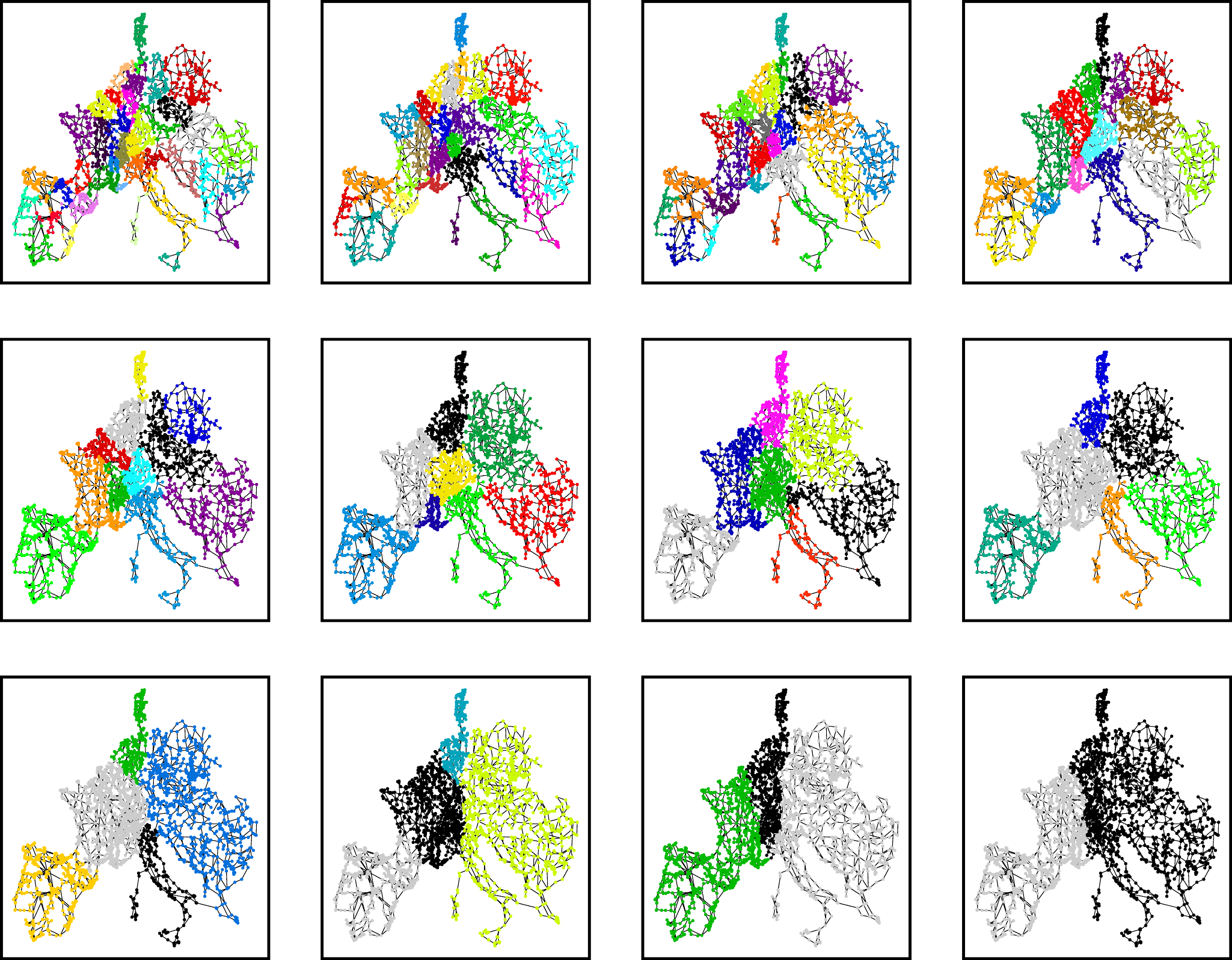}
 \caption{\textbf{Multiscale community structure of the European power grid with stability}. The illustrative partitions shown correspond to
Markov times $t=\{ 0.81, 2.12, 2.62, 11.75, 20.08, 34.29, 94.79, 173.49,
389.77, 1176.77, 1919.14, 10000\}$ (from left to right, top to bottom) and have been selected based on their relative robustness. The dynamic zooming provided by the Markov time provides a progressively coarser representation of the network that captures geopolitical and commercial features of the grid. }
 \label{fig:6}
\end{figure*}

Through its intrinsic dynamic sweeping, stability reveals a multiscale structure with meaningful subregions of different sizes at various Markov times. The Markov times highlighted in Figure~\ref{fig:5} have been selected according to the relative decrease in variation of information, which indicates a more robust partitioning.
Interestingly, the communities found by stability at different times appear to be related to historical and commercial features of the power grid network.
For larger times, the coarse communities correspond well with big historical monopolies, basically identified with nations. Germany provides an exception to this scheme since the German power grid is split between four large companies (with one covering the eastern part of Germany)\footnote{See \url{http://de.wikipedia.org/wiki/Stromnetz\#Netzbetreiber} }. This is reflected faithfully by the community structure detected by stability (Fig.~\ref{fig:5}D). For shorter times (Fig.~\ref{fig:5}B-C), we get communities on a sub-national scale that also correspond to regional operators, e.g., France gets split up into several communities which overlap well with the regional organization of the French power grid~\footnote{For a map of the French regional electrical companies see \url{http://www.rte-france.com/fr/nous-connaitre/qui-sommes-nous/organisation-et-gouvernance/le-siege-et-les-unites-regionales}}. Similar effects are observed in Spain, Italy and Switzerland.   In Figure~\ref{fig:6}, we show a representation of the communities found as the dynamic zooming lens of the Markov process is applied to this network. The detailed analysis of these observations will be the object of future work.

The dynamic zooming provided by the Markov time is not only the key ingredient to detect communities without a restriction of scales but also the robustness (or lack of robustness) of specific communities is revealing of characteristics of those communities. In the case of the grid,  interesting effects can be found when looking at the way communities coarsen as the Markov time gets larger. For instance, the communities in Switzerland appear to flip between different adjacent communities as the Markov time increases. This lack of robustness indicates a strong shared interconnectedness of the Swiss communities with its neighboring groups. This aligns well with the known fact that the
Swiss power-grid is an important mediating hub in the center of Europe with $11 \%$ of Europe's electricity flowing through Switzerland~\footnote{See \url{https://www.swissgrid.ch/swissgrid/en/home/europa/european_overview/ch_in_europe.html}}.

\section*{Discussion}

The examples considered above illustrate how one-step methods, which are tuned to detect clique-like or low diameter communities, are prone to displaying a form of overpartitioning for networks with non clique-like community structure. This is the result of the existence of a field-of-view limit for structural, one-step algorithms, such as Infomap and modularity. The field-of-view limit establishes an upper limit on the size of the communities that can be detected, as given by their effective intra-community distance.
Therefore, non clique-like communities present a challenge because they have an inherent scale that might fall beyond the field-of-view of standard one-step methods. When the intra-community effective distance is larger than the field-of-view, this leads to overpartitioning.
The field-of-view  appears on the opposite scale of the well-known resolution limit~\cite{Fortunato2007}, which indicates a lower limit size below which modularity cannot detect communities.

Indeed, Infomap can be seen as having being designed to resolve absolutely the fine scale (hence no resolution limit) by optimizing for locally clique-like substructures. This feature, which makes it extremely successful in the analysis of networks with high density of connections, makes it `myopic' with respect to larger non clique-like structures. Therefore, Infomap suffers from a large field-of-view limit, even more acutely than modularity, as shown in the examples above.  Interestingly, the fact that Infomap can also be seen as a one-step method from a dynamical perspective, as discussed above, has allowed us to propose a multi-step correction of the Map formalism that may provide a remedy for this problem~\cite{Schaub2011}.
Modularity has an intrinsic scale dependent on the overall size (weight) of the graph which allows it to resolve communities in a range between the resolution and field-of-view limits. Depending on the particular graph, this scale may be well matched to the community structure present (in which case, modularity works) or may be too large (underpartitioning, affected by the resolution limit) or too small (overpartitioning, affected by the field-of-view limit).
Stability on the other hand applies a dynamic zooming across all scales following the Markov time evolution and thus does not impose \textit{a priori} a specific scale to detect the community structure. It is important to remark that the dynamic zooming is a key aspect of the analysis: only by scanning through Markov time one can reveal whether a specific (time-)scale is meaningful for the problem at hand or whether it merely corresponds to an artificial, non-robust partition.

Although the standard assumption of communities as probabilistic clique-like groups is well motivated and relevant for a number of important complex systems, our examples highlight the fact that this view of community is not always representative of the structures found in networks of current interest.  Segments in an image, functional subunits of a protein or geographic entities in the power network are modular substructures of networks that are only adequately identified when multi-step community detection methods, such as stability optimization, are employed. Furthermore, multi-step methods such as stability can also be used for the analysis of clique-like community structures, as shown in Ref.~\cite{Traag2011} with benchmark models (see Fig.~2 therein and note that the Reichardt-Bornholdt Potts method\cite{Reichardt2004, Reichardt2006} has been shown~\cite{Delvenne2010,Lambiotte2009} to correspond to a linearization of our partition stability formalism). As recent work suggests \cite{Lancichinetti2011a}, however, it may not be possible to resolve highly inhomogeneous community structure at a single (fixed) time, and further research needs to be pursued to develop local methods that do not infer a set of scales from the global graph.

Arguably, communities that are non clique-like are in some instances most relevant, in the sense that they correspond to subsystems in which all parts are related but not necessarily in a direct manner. This is a specific network viewpoint for data analysis, as opposed to a generic set of relationships between elements based on correlation.
In fact, the analysis of networks with clique-like communities, corresponding to subsystems where all nodes interact with each other, may be more appropriately pursued through multivariate statistics, rather than explicit network analysis.
This realization also has implications for current efforts to construct networks from data where, in contrast to correlation matrices, locally sparser networks are favored as a means to reveal the underlying systemic connections of the data~\cite{Bullmore2009}.

Nearly all commonly used benchmark models to date have adopted a clique-like notion of community, a fact that needs to be taken into account when considering comparative tests performed on these benchmarks, which are likely to be favorable
towards methods that have been designed  for the detection of a clique-like notion of community~\cite{Lancichinetti2009}.
Although these benchmark graphs reflect adequately the community structure of many important networks and datasets, we have shown that many real-world graphs are likely to have non-clique community substructures. This observation hints at the need for a broader set of benchmarks (e.g., random geometric graphs) that go beyond a block-averaged structure and can reflect the specific properties of networks from different application areas.

No community detection method will be universally optimal for the analysis of all networks and, as pointed above, trade-offs between specific and generic features are to be expected. Our findings highlight the critical importance of a detailed assessment of the assumed definition of community and of its appropriateness to the features of the network to be analyzed. Our work also reinforces the need for a careful consideration of the scales that might be implicit in community detection algorithms that can lead to the identification of non-relevant structures as communities. Our use of the multi-step stability framework proposes to circumvent the implicit assumption of a scale by considering community detection as a zooming process through all scales in a systematic manner, as given by a Markov diffusion on the graph.  Alternatively, one can understand the stability formalism as finding the particular time scale at which modules in the graph will be seen as clique-like communities after the process diffuses on the graph.

However, and specifically in the case of non-regular graphs, there remain issues inherent to the fact that all the methods considered in this paper use global definitions of community. As stated above, there is a need for further research towards local methods for community detection that avoid the restrictions imposed by the global graph scale and potentially allow for a soft, overlapping partitioning of the network. More generally, although a huge set of community detection algorithms has been presented to date, it is still an open problem to establish connections between the underlying themes (conceptual and algorithmic) that these methods implement and the applications for which they might be especially suitable. This will be the object of future work.

\section*{Acknowledgments}
The authors thank Renaud Lambiotte for fruitful discussions; Arnaud Browet for kindly sharing code to create the graphs from images; and Antoine Delmotte for discussions and for help with the creation of the figures and graphs of the protein examples. We are thankful to Mart\'{\i} Rosas-Casals for making the power grid data available.
M.T.S. has been supported by a grant from the EPSRC of the UK under the \textit{Mathematics underpinning the Digital Economy} program (to S.N.Y. and M.B.) and by a grant of the U.S. Office of Naval Research (to S.N.Y.). J.-C. D. is supported by the Belgian Programme of Interuniversity Attraction Poles and an Action de Recherche Concert\'{e}e (ARC) of the French Community of Belgium.

\bibliography{non_clique}

\section*{Figure Legends}

\end{document}